# Non-Uniform Sampling Reconstruction for Symmetrical NMR Spectroscopy by Exploiting Inherent Symmetry

Enping Lin[a], Ze Fang[a], Yuqing Huang[a], Yu Yang[a,*], and Zhong Chen[a,*]

[a]*Department of Electronic Science, State Key Laboratory of Physical Chemistry of Solid Surfaces, Xiamen University, Xiamen, Fujian 361005, China*

---

**Abstract**

Symmetrical NMR spectroscopy constitutes a vital branch of multidimensional NMR spectroscopy, providing a powerful tool for the structural elucidation of biological macromolecules. Non-Uniform Sampling (NUS) serves as an effective strategy for averting the prohibitive acquisition time of multidimensional NMR spectroscopy by only sampling a few points according to NUS sampling schedules and reconstructing missing points via algorithms. However, current sampling schedules are unable to maintain the accurate recovery of cross peaks that are weak but important. In this work, we propose a novel sampling schedule termed as SCPG (Symmetrical Copy Poisson Gap) and employ CS (Compressed Sensing) methods for reconstruction. We theoretically prove that the symmetrical constraint, apart from sparsity, is implicitly implemented when SCPG is combined with CS methods. The simulated and experimental data substantiate the advantage of SCPG over state-of-the-art 2D Woven PG in the NUS reconstruction of symmetrical NMR spectroscopy.

---

* Corresponding author.

*E-mail address: chenz@xmu.edu.cn (Z. Chen), yuyang15@xmu.edu.cn (Y. Yang).*

## 1. Introduction

NMR spectroscopy offers a robust and non-invasive tool for applications in various fields, such as chemistry, biology, and materials science[1-4]. Symmetrical NMR spectroscopy is an essential branch of multidimensional NMR spectroscopy possessing symmetrical structure, such as COrrelated SpectroscopY (COSY)[5], Nuclear Overhauser Enhancement SpectroscopY (NOESY)[6], Total Correlation Spectroscopy (TOCSY)[7], solid-state NMR 13C-13C correlation based Dipolar Assisted Rotational Resonance (DARR)[8] or Proton-Driven Spin Diffusion (PDSD)[9] and higher dimensional spectroscopy encompassing these symmetrical modules. These symmetrical spectroscopy techniques provide a powerful tool for structural elucidation and functional analysis of biological macromolecules, e.g., peptides, proteins, and glycogen. However, multidimensional NMR experiments suffer substantial experimental time, which can be significantly reduced by the Non-Uniform Sampling (NUS) strategy in which only a few data points are sampled. For NUS NMR data, algorithms are employed to reconstruct the desired spectra. The quality of reconstructed NUS NMR spectra is mainly determined by the sampling schedule and the reconstruction algorithm.

Sampling schedules aim at selecting a set of optimal data points representing the whole spectrum with as much information as possible. Sampling schedules are normally based on a probabilistic model. The most straightforward schedule is Random sampling, in which every data point is sampled with uniform probability. However, the Random sampling schedule is not always an optimal choice since the data points of the free induction decay (FID) signals might not be equally important. Considering the different contributions of different data points in FID, some weighted probabilistic sampling schedules have been proposed, e.g., exponential weight sampling schedule[10-12] and others[13-18]. Instead of treating the sampling points as a probabilistic variable, Poisson Gap (PG) sampling schedule[19] regards the gap of two neighboring sampling points as a probabilistic variable that follows the sinusoid-weighted Poisson distribution, thus leading to a tight sampling in the starting and ending of FID and a loose sampling in the middle part. PG sampling schedule has been generally applied to NUS NMR experiments and recognized as one of the most effective sampling schedules. The recent literature[20] has shed some light on the theoretical relationship between the sampling schedule and NMR spectrum reconstruction results. For multidimensional NUS experiments, PG sampling schedule is extended into nD Woven PG sampling schedule[21, 22].

Reconstruction algorithms aim to retrieve the whole spectra based on the incomplete set of sampled FID data points. A variety of reconstruction techniques, such as multidimensional decomposition[23], direct FT of zero-augmented data[24], maximum entropy[25, 26], and Compressed Sensing (CS)[27-30], can be employed to retrieve the spectra from the reduced sampling data points. CS methods are generally regarded as the most effective technique, supported by the theory that strictly sparse signals can be reconstructed perfectly from a set of sampling points with a size in the order of $r\ log(n/r)$, where $n$ is the total grid size and $r$ denotes the number of non-zero points[31-34]. Correspondingly, by constraining the sparsity of NMR spectroscopy, CS methods can effectively reconstruct NMR spectroscopy from substantially reduced data points. CS methods are comprehensively and systematically reviewed in[27] and are implemented mainly based on two approaches, i.e., regularized inversion and greedy algorithms. The former treats the reconstruction as the $l_p$-norm $(0 \leq p \leq 1)$ regularized inverse problems, while the latter iteratively selects the non-zero values to fit the NUS data.

CS methods tend to maintain the strong peaks and omit or suppress weak peaks to approach spectral sparsity. The symmetrical NMR spectroscopy possesses a wide dynamic range of spectral peaks— strong peaks normally lie in diagonal positions (diagonal peaks), while weak ones symmetrically lie in off-diagonal positions (cross peaks). Thus, cross peaks are of insufficient reconstruction quality when the NUS rate is low, even when CS methods combined with state-of-the-art 2D Woven PG sampling are employed. In this work, on the basis of CS methods, we exclusively exploit the inherent symmetry, apart from sparsity, for improving the recovery of cross peaks of symmetrical NMR spectroscopy. The symmetrical constraint is implicitly imposed by the proposed Symmetrical Copy Poisson Gap (SCPG) sampling schedule. We theoretically prove that, with the SCPG sampling schedule, the spectrum reconstructed by CS methods would be strictly symmetric. Simulated and experimental data reconstructions corroborate that CS methods combined with the SCPG sampling schedule would substantially improve the reconstruction quality of symmetrical NMR spectroscopy compared to state-of-the-art 2D Woven PG sampling. The rest of the paper is arranged as follows. In Section 2, we prepare the preliminary knowledge for SCPG; in Section 3, we detail SCPG and its characteristics when CS is combined; in Section 4, simulated and experimental data are employed to validate the advantage of SCPG; Section 4 is the conclusion of this work.

Throughout this paper, scalars are written as lowercase letters; vectors or arrays as bold lowercase letters; matrices or operators as bold capital letters; functions as italic lowercase letters. The superscripts '$-1$', '$T$', '$*$' and '$H$' indicate the inversion, transposition, conjugate and conjugate transposition of the corresponding vectors or matrices, respectively. '$\odot$' is the element-wise product sign. Additionally, adjoint operators are written as operator signs with a superscript '$H$'; $\boldsymbol{I}$ denotes the identity matrix; $\boldsymbol{F}$ denote the Fourier Transform (FT) matrix, $\boldsymbol{F}^H$ or $\boldsymbol{F}^{-1}$ is the Inverse Fourier Transform (IFT) matrix, which satisfies $\boldsymbol{F}\boldsymbol{F}^H = \boldsymbol{F}^H\boldsymbol{F} = \boldsymbol{I}$, i.e., $\boldsymbol{F}^H = \boldsymbol{F}^{-1}$. $\boldsymbol{F}_{2D}$ and $\boldsymbol{F}_{2D}{}^{-1}$ denote 2D FT and IFT operators, respectively. For a square matrix $\boldsymbol{X}$, we have $\boldsymbol{F}_{2D}\boldsymbol{X} = \boldsymbol{F}\boldsymbol{X}\boldsymbol{F}^T$ and $\boldsymbol{F}_{2D}{}^{-1}\boldsymbol{X} = \boldsymbol{F}^H\boldsymbol{X}(\boldsymbol{F}^H)^T$. $\|.\|_p$ denotes the $l_p$-norm ($0 \leq p \leq 1$) of a vector and $\|\boldsymbol{x}\|_p$ is defined as $(\sum_i |x_i|^p)^{1/p}$ where $|x_i|$ is the absolute value of the $i$-th entry of $\boldsymbol{x}$. If $\boldsymbol{x}$ is a matrix or a higher dimensional array, it should be vectorized in advance. $abs(.)$ denotes the absolute function. $sign(.)$ denotes the sign function. $\mathbb{C}^{n_1 \times n_2 \cdots \times n_m}$ denotes the set consisting of all $m$-dimension complex-valued arrays, the size of which is detailed in the superscripts; $\langle .,. \rangle$ is the sign of inner product. All computations in this work are carried out in MATLAB® 2021b in a laptop with ADM Ryzen 7 5800H 16-core CPU and 16GB memory.

## 2. Preliminary

### *2.1. Poisson Gap (PG) for NUS NMR Sampling*

If a random variable $\mathcal{X}$ is allowed to be non-negative integers and has the probability of $\mathcal{X} = k$ ($k = 0,1,2, \ldots$) given as

$$\mathrm{P}(\mathcal{X} = k) = \alpha^k \exp(-\alpha)/k!, \alpha > 0, \tag{1}$$

then, $\mathcal{X}$ follows the Poisson distribution with a parameter $\alpha$ and denoted as $\mathcal{X} \sim \text{Pois}(\alpha)$. The mathematical expectation $\text{E}(\mathcal{X})$ and variation $\text{Var}(\mathcal{X})$ are both $\alpha$. In PG sampling schedule, the gaps between sampling points obey the Poisson distribution. In the original work of the application of PG sampling schedule in NUS NMR[19], Hyberts et al. state that the initial and final parts of the FID signals should be sampled with smaller gaps. Thus, the average gap length $\alpha$ is modulated with a sinusoidal variation, i.e.,

$$\alpha = \gamma \sin(\theta n / n_{max}), \tag{2}$$

with $\theta = \pi$. In (2), $n$ is the index of evolution time grids; $n_{max}$ is the length of evolution time grids; $\gamma$ is an adjusting constant for the desired sampling density. It is noteworthy that if the final part of FID signals is highly perturbed due to the decay and apodization operation, it can be sampled with larger gaps by setting $\theta = \pi/2$.

For simplicity, in this work, without further specification, PG indicates the PG modulated with $0 \sim \pi$ sinusoidal variation. For 2D NUS NMR experiments, Hyberts et al. further proposed a 2D Woven PG sampling schedule, where the 2D plane is weaved with 1D PG lines, as illustrated in Figure 1a[21, 22]. However, what Figure 1a does not reveal is that the lines have to be shifted backward until the beginning is occupied. In addition, $n$ and $n_{max}$ in (2) are reformulated as $n = n_1 + n_2$ and $n_{max} = n_{max}^{(1)} + n_{max}^{(2)}$, where $n_i$ and $n_{max}^{(i)}$ $(i = 1,2)$ denote the index of time points and the total number of time grids along the $i$-th dimension. Figure 1b and 1c show two examples modulated with $0 \sim \pi/2$ and $0 \sim \pi$ sinusoidal variation, respectively.

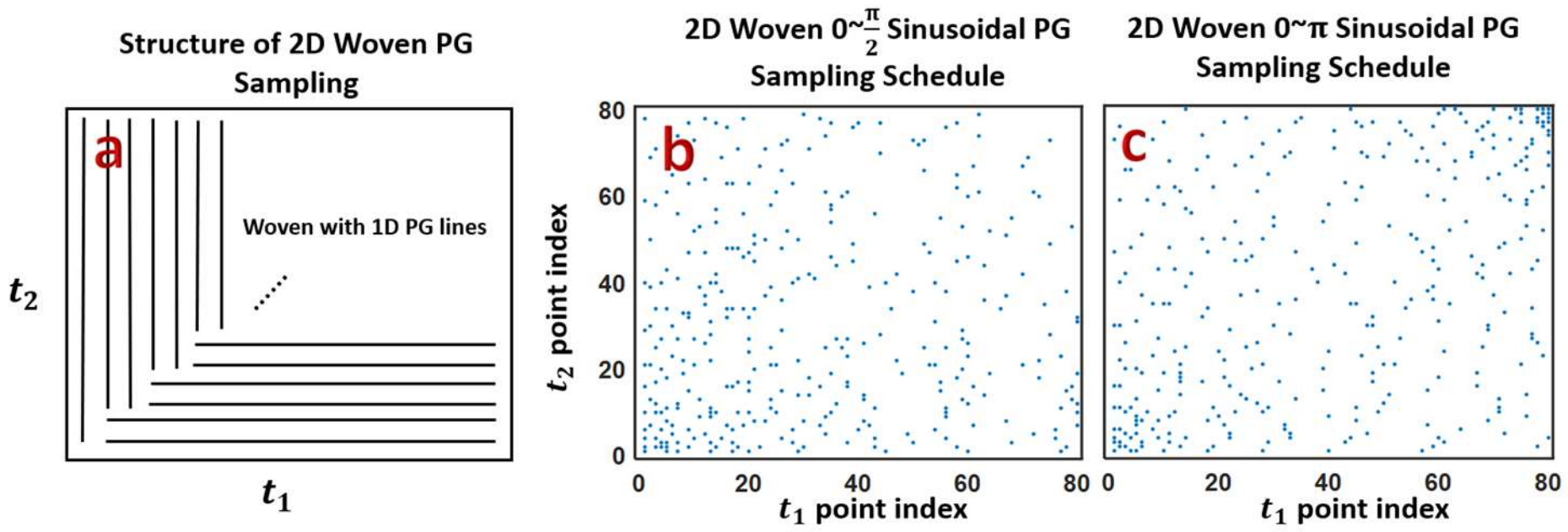

**Figure 1.** Illustration of 2D Woven PG sampling. (a) The structure of 2D Woven PG sampling. instances of 2D Woven PG sampling schedule modulated with $0\sim\pi/2$ and $0\sim\pi$ are shown in (b) and (c), respectively. Blue dots indicate the sampled data points.

*2.2. CS methods for NUS NMR Reconstruction*

For 1D NUS NMR reconstruction, CS methods attempt to solve the following mathematical problem to reconstruct a matched and sparse spectrum:

$$\arg\min_{\boldsymbol{x}} \|\boldsymbol{x}\|_p, \ s.t. \|\boldsymbol{P}_\Omega \boldsymbol{F}^{-1}\boldsymbol{x} - \boldsymbol{y}\|_2 \leq \varepsilon, \tag{3}$$

where $\boldsymbol{x}$ is the to-be-reconstructed NMR spectrum; $\boldsymbol{y}$ is the NUS FID data; $\boldsymbol{F}^{-1}$ is the inverse Fourier Transform matrix; $\Omega$ is the index set of sampling points; $\boldsymbol{P}_\Omega$ is the NUS operator; $\|\boldsymbol{x}\|_p$ measures the sparsity of $\boldsymbol{x}$. $\|\boldsymbol{P}_\Omega \boldsymbol{F}^{-1}\boldsymbol{x} - \boldsymbol{y}\|_2$ measures the data consistency of the reconstructed spectrum $\mathbf{x}$ and NUS sampled FID data $\boldsymbol{y}$; $\varepsilon$ denotes the estimated noise level. Equation (3) could be converted to the following $l_p$-norm regularized inverse problem:

$$\arg\min_{\boldsymbol{x}} \|\boldsymbol{P}_\Omega \boldsymbol{F}^{-1}\boldsymbol{x} - \boldsymbol{y}\|_2^2/2 + \lambda\|\boldsymbol{x}\|_p, \tag{4}$$

where $\lambda$ is the regularization parameter. For a certain $\lambda$, the optimal solution of (4) is also the solution of (3) for some $\varepsilon$, and vice versus. Candès et al. have shown that the $l_1$-norm regularization allows us to find the sparsest spectrum that matches the NUS FID data $\mathbf{y}$[31], which works equally well for $p < 1$[34]. A host of optimization algorithms were proposed for solving $l_1$-norm regularized inverse problem (i.e., $p = 1$ for (4)), which is the most commonly used and effective formulation in CS area due to its convexity.

On the other hand, (3) can be directly solved by greedy algorithms, the most renowned one of which is Iterative Soft Thresholding (IST) for $p = 1$ in (3). In this work, we only discuss the common and effective variant of IST, which is named IST-D by Sun et al. [35]. The basic IST-D framework includes six steps:

1. NUS FID signals with non-sampled points filled with zeros are treated as the input;

2. Perform FT on the input FID and truncate the spectrum with a threshold;

3. Parts of the spectrum above the thresholding line are added to the temporary spectrum, which will be the output spectrum when iterations are over;

4. Perform IFT on the current temporary spectrum and obtain the corresponding FID signal;

5. Zero the non-sampled parts of the obtained FID, and go to step 6;

6. Subtract the obtained zero-filling FID from the zero-filling NUS FID signal, perform FT on the resulting FID and obtain the corresponding spectrum, then treat the spectrum as an input of step 2.

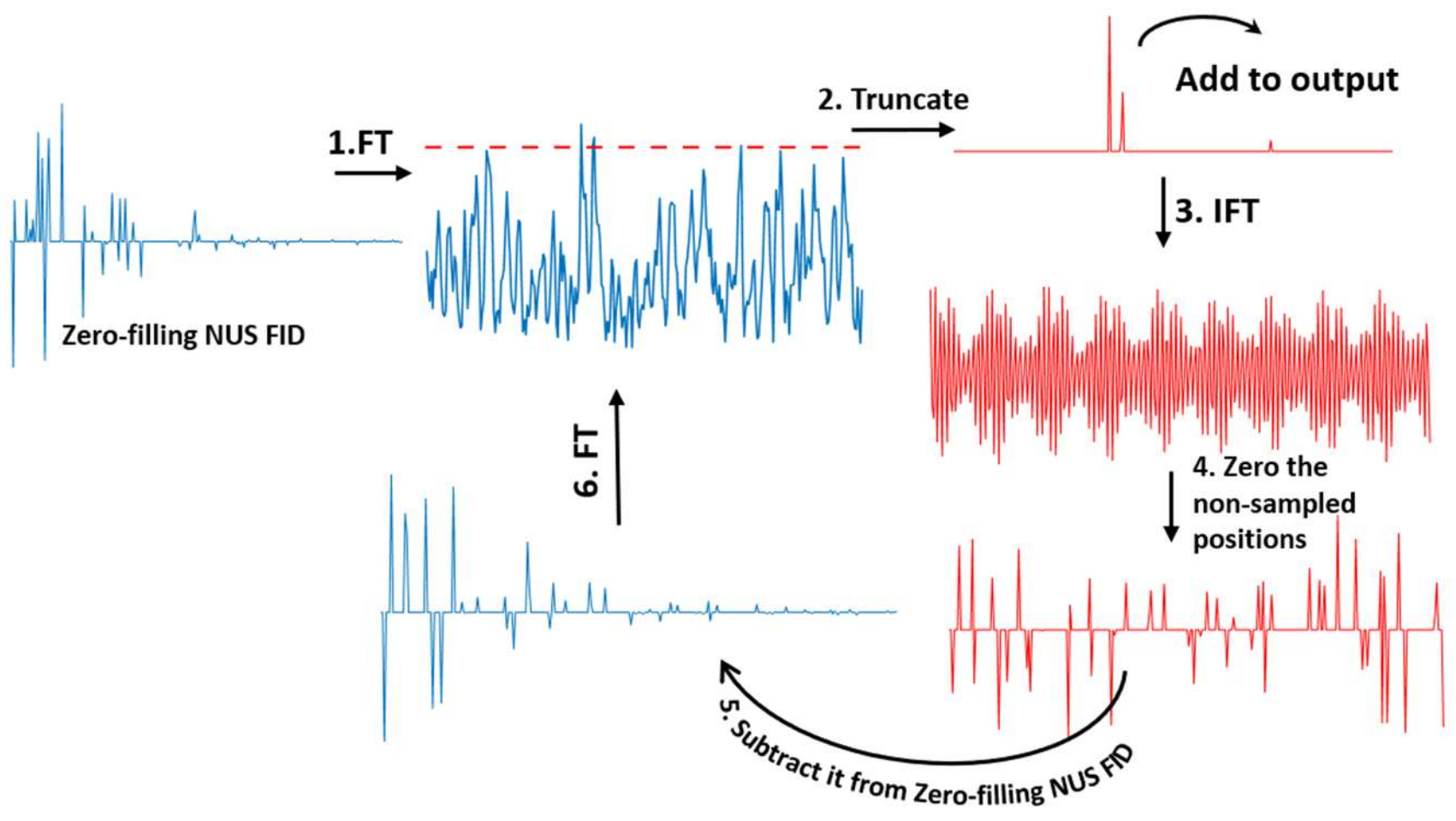

**Figure 2**. The illustration of six-step frameworks of IST-D.

Sparsity is enhanced as the iteration of IST-D increases. The iteration terminates if either of the two conditions below is satisfied: 1. The residue between the reconstructed FID signal and the given NUS FID signal is less than the required tolerance; 2. The maximum iteration is achieved. Many effective NUS reconstruction methods or commercial software belong to IST-D, e.g., Implementation of IST at Harvard Medical School (istHMS) [22], Sparse multidimensional iterative lineshape-enhanced (SMILE) reconstruction[36, 37], and MddNMR software[23, 38]. These IST-D-based methods or software further contain some specific considerations for performance improvement and automatic setup of parameters. For simplicity and generality, the discussion in this work is only concerned with the basic IST-D frameworks.

## 3. Theory

### *3.1. Symmetry*

Symmetry is a common occurrence in nature, indicating a sense of simplicity since only half of signal is able to represent the whole. A square matrix has four symmetry axes, i.e., central horizontal, central vertical, diagonal and anti-diagonal lines, leading to four kinds of symmetries. According to the symmetry axis, any square matrix $\boldsymbol{X}$ can be divided into three parts: the entries located on the symmetry axis, the real part, and the imaginary part. The real and imaginary parts are relative to each other.

Define a symmetry-permutation operator $\mathcal{S}$, and $\mathcal{S}(\boldsymbol{X})$ operation mirror flips the real and imaginary parts of $\boldsymbol{X}$ while preserving the elements along the symmetry axis. Then a square matrix $\boldsymbol{X}$ is symmetrical if

$$\boldsymbol{X} = \mathcal{S}(\boldsymbol{X}). \tag{5}$$

***Definition*** 1: We say a function $\mathcal{R}(\boldsymbol{X}): \mathbb{C}^{n\times n} \to \mathbb{C}$ is symmetry-invariant if $\mathcal{R}(\boldsymbol{X}) = \mathcal{R}\big(\mathcal{S}(\boldsymbol{X})\big)$; We say an operator $\boldsymbol{A}$ is symmetry-invariant if $\boldsymbol{A}\mathcal{S}(\boldsymbol{X}) = \mathcal{S}(\boldsymbol{A}\boldsymbol{X})$.

It is easy to infer that $\mathcal{S}(\boldsymbol{A}\boldsymbol{X}) = \boldsymbol{A}\boldsymbol{X}$ if $\boldsymbol{X}$ is symmetrical and $\boldsymbol{A}$ is symmetry-invariant, and that $\boldsymbol{B} = \boldsymbol{B}_1\boldsymbol{B}_2 \ldots \boldsymbol{B}_n$ is symmetry-invariant if $\boldsymbol{B}_1, \boldsymbol{B}_2, \ldots, \boldsymbol{B}_n$ are all symmetry-invariant.

Consider a regularized inverse problem

$$\arg\min_{\mathbf{X}} \frac{1}{2} ||\mathbf{AX} - \mathbf{Y}||_2^2 + \lambda\mathcal{R}(\mathbf{X}), \tag{6}$$

where $\boldsymbol{A}$ is a linear operator, $\boldsymbol{Y}$ is an input symmetrical square matrix, $\boldsymbol{X}$ is an output square matrix, $\mathcal{R}(.)$ is a norm function, $\lambda$ is a non-negative valued regularization parameter trading off the balance of fidelity and regularization terms.

***Theorem*** 1: If the linear kernel operator $\boldsymbol{A}$ and norm function $\mathcal{R}(.)$ are symmetry-invariant, then at least one of the optimal solutions of (6), denoted as $\boldsymbol{X}^*$, is symmetrical, i.e., $\boldsymbol{X}^* = \mathcal{S}(\boldsymbol{X}^*)$.

Proof: For any asymmetrical matrix $\widetilde{\boldsymbol{X}}$, we can construct a symmetrical matrix $\overline{\boldsymbol{X}} = \left(\widetilde{\boldsymbol{X}} + \mathcal{S}(\widetilde{\boldsymbol{X}})\right)/2$. According to the triangular inequality of the norm, there is: $\|\boldsymbol{A}\overline{\boldsymbol{X}} - \boldsymbol{Y}\|_2 \leq \left[\|\boldsymbol{A}\widetilde{\boldsymbol{X}} - \boldsymbol{Y}\|_2 + \|\boldsymbol{A}\mathcal{S}(\widetilde{\boldsymbol{X}}) - \boldsymbol{Y}\|_2\right]/2 = \|\boldsymbol{A}\widetilde{\boldsymbol{X}} - \boldsymbol{Y}\|_2$. Notice that $\boldsymbol{A}\mathcal{S}(\widetilde{\boldsymbol{X}}) = \boldsymbol{A}\widetilde{\boldsymbol{X}}$ holds for symmetry-invariant operator $\boldsymbol{A}$. Thus, we conclude

$$\|\boldsymbol{A}\overline{\boldsymbol{X}} - \boldsymbol{Y}\|_2^2/2 \leq \|\boldsymbol{A}\widetilde{\boldsymbol{X}} - \boldsymbol{Y}\|_2^2/2. \tag{7}$$

Similarly, we have

$$\lambda\mathcal{R}(\overline{\boldsymbol{X}}) = \lambda\mathcal{R}\left(\widetilde{\boldsymbol{X}} + \mathcal{S}(\widetilde{\boldsymbol{X}})\right)/2 \leq \lambda\left(\mathcal{R}(\widetilde{\boldsymbol{X}}) + \mathcal{R}\left(\mathcal{S}(\widetilde{\boldsymbol{X}})\right)\right)/2 = \lambda\mathcal{R}(\widetilde{\boldsymbol{X}}). \tag{8}$$

Summing up the above two inequalities, we have

$$\|\boldsymbol{A}\overline{\boldsymbol{X}} - \boldsymbol{Y}\|_2^2/2 + \lambda\mathcal{R}(\overline{\boldsymbol{X}}) \leq \|\boldsymbol{A}\widetilde{\boldsymbol{X}} - \boldsymbol{Y}\|_2^2/2 + \lambda\mathcal{R}(\widetilde{\boldsymbol{X}}). \tag{9}$$

(9) reveals that for any asymmetrical matrix $\widetilde{\boldsymbol{X}}$, we can always construct a corresponding symmetric matrix $\overline{\boldsymbol{X}}$, which has an equal or lower objective function value compared to $\widetilde{\boldsymbol{X}}$. Thus, we conclude at least one of the optima of (6) $\boldsymbol{X}^*$ is symmetrical, i.e., $\boldsymbol{X}^* = \mathcal{S}(\boldsymbol{X}^*)$.

Many commonly-applied regularization functions are symmetry-invariant, e.g., $l_1$-norm for sparsity constraint, nuclear norm for low-rank constraint, and $l_2$-norm for smoothness constraint. The inequality (7) indicates that there is no conflict between symmetry and sparsity, low-rank, smoothness and other possible prior knowledge, i.e., we can always find a corresponding symmetrical matrix without loss of degree of sparsity, low-rank, smoothness and etc. Furthermore, *Theorem* 1 reveals the solution of (6) must be symmetrical if the solution to (6) is unique, which can be guaranteed when $\boldsymbol{A}$ satisfies the Restricted Isometry Property (RIP) condition[34].

*3.2. Some Useful Symmetry-Invariant Operators*

In this work, without further specification, symmetry refers to diagonal-symmetry and $\mathcal{S}$ is the transposition of a matrix under this condition. Below, we introduce some useful symmetry-invariant linear operator for a square matrix and give related proofs.

***Lemma*** 1: 2D FT and IFT operators, $\boldsymbol{F}_{2\mathrm{D}}$ and ${\boldsymbol{F}_{2\mathrm{D}}}^{-1}$, are symmetry-invariant.

Proof: For 2D FT operator, we have $\boldsymbol{F}_{2\mathrm{D}}\boldsymbol{X}^T = \boldsymbol{F}\boldsymbol{X}^T\boldsymbol{F}^T = (\boldsymbol{F}\boldsymbol{X}\boldsymbol{F}^T)^T = (\boldsymbol{F}_{2\mathrm{D}}\boldsymbol{X})^T$. For 2D IFT operator, we have ${\boldsymbol{F}_{2\mathrm{D}}}^{-1}\boldsymbol{X}^T = \boldsymbol{F}^H\boldsymbol{X}^T\boldsymbol{F}^* = (\boldsymbol{F}^H\boldsymbol{X}\boldsymbol{F}^*)^T = \left({\boldsymbol{F}_{2\mathrm{D}}}^{-1}\boldsymbol{X}\right)^T$.

The FID signal of 2D symmetrical NMR spectrum is also symmetrical, which is corollary of *Lemma* 1.

***Lemma*** 2: The soft thresholding operator with $\beta \geq 0$ of a matrix $\boldsymbol{X}$, $\mathrm{SHR}_\beta(\boldsymbol{X})$, defined as:

$$\mathrm{SHR}_\beta(\boldsymbol{X}) = (sign(\boldsymbol{X})/2)\odot\big(abs(\boldsymbol{X}) - \beta + abs(abs(\boldsymbol{X}) - \beta)\big), \tag{10}$$

is symmetry-invariant.

Proof: Based on the basic fact that $sign(\boldsymbol{X})^T = sign(\boldsymbol{X}^T)$, $abs(\boldsymbol{X})^T = abs(\boldsymbol{X}^T)$ and $(\boldsymbol{X} - \beta)^T = \boldsymbol{X}^T - \beta$, we have

$$\begin{aligned}\mathrm{SHR}_\beta(\boldsymbol{X}^T) &= (sign(\boldsymbol{X}^T)/2)\odot\big(abs(\boldsymbol{X}^T) - \beta + abs(abs(\boldsymbol{X}^T) - \beta)\big)\\ &= (sign(\boldsymbol{X})^T/2)\odot\big(abs(\boldsymbol{X})^T - \beta + abs(abs(\boldsymbol{X})^T - \beta)\big).\\ &= \mathrm{SHR}_\beta(\boldsymbol{X})^T\end{aligned} \tag{11}$$

***Definition*** 2: Given a matrix $\boldsymbol{X}$, we define a NUS operator $\boldsymbol{P}_\Omega$ where $\Omega$ is the index set of sampling points. When $\boldsymbol{P}_\Omega$ is performed on $\boldsymbol{X}$, $\boldsymbol{P}_\Omega\boldsymbol{X}$ fetches the data points of $\boldsymbol{X}$ according to the index set $\Omega$ and vectorizes the fetched data points as the output. The adjoint operator of $\boldsymbol{P}_\Omega$ is written as ${\boldsymbol{P}_\Omega}^H$. $\boldsymbol{P}_\Omega$ and ${\boldsymbol{P}_\Omega}^H$ comply with the following identity (12), which describes the relationship of a bounded linear operator and its adjoint operator.

$$\langle\boldsymbol{P}_\Omega\boldsymbol{X}, \boldsymbol{y}\rangle \equiv \langle\boldsymbol{X}, {\boldsymbol{P}_\Omega}^H\boldsymbol{y}\rangle, \forall\, \boldsymbol{X} \in \mathbb{C}^{N\times N},\ \boldsymbol{y} \in \mathbb{C}^{r\times 1}\ , \tag{12}$$

where $r$ is the cardinality of $\Omega$.

As illustrated with a toy example in Figure 3, ${\boldsymbol{P}_\Omega}^H\boldsymbol{y}$ returns a matrix of the size of $\boldsymbol{X}$, whose elements indexed by $\Omega$ are filled with the corresponding elements of $\boldsymbol{y}$ and otherwise zero. Obviously, ${\boldsymbol{P}_\Omega}^H\boldsymbol{P}_\Omega\boldsymbol{X}$ returns a matrix of the size of $\boldsymbol{X}$ with elements indexed by $\Omega$ the same as those of $\boldsymbol{X}$ and the others as zero. The following lemma can be easily derived:

***Lemma*** 3: If $\Omega$ consists of symmetrical index pair, ${\boldsymbol{P}_\Omega}^H\boldsymbol{P}_\Omega$ is symmetry-invariant.

## 4. Method

### *4.1. Symmetrical Copy Poisson Gap (SCPG)*

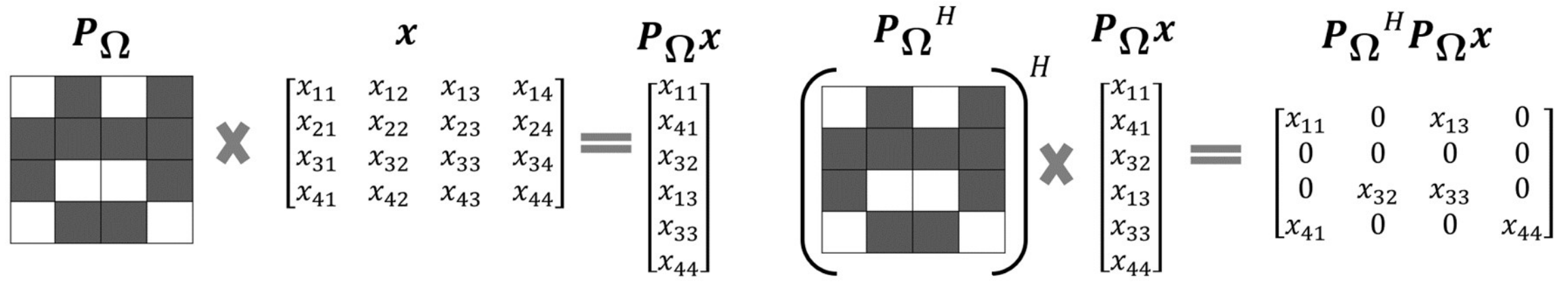

**Figure 3**. Illustration of $\boldsymbol{P}_\Omega$ and $\boldsymbol{P}_\Omega{}^H$ operators performed on a given matrix $\boldsymbol{X}$. In the grid table, white grids indicate the sampled points, while gray ones indicate the non-sampled points.

To better reconstruct a symmetrical spectrum, we propose Symmetrical Copy Poisson Gap (SCPG) sampling schedule to construct a pseudo-symmetrical NUS operator $\boldsymbol{P}_\Omega$. Here, we introduce a term, symmetrical pair, to describe this pseudo-symmetrical sampling schedule.

***Definition*** 3: For 2D symmetrical NMR FID signal $f \in \mathbb{C}^{N\times N}$, $f(i,j)$ and $f(j,i)$ $(1 \leq i \leq j \leq \mathrm{N})$ constitute a symmetrical pair denoted as $sp(k)$ where $k = j(j-1)/2 + i$. If $j \neq i$, i.e., $f(i,j)$ is an off-diagonal point, the symmetrical pair $sp(k)$ has two points. If $j = i$, i.e., $f(i,j)$ is a diagonal point, the symmetrical pair $sp(k)$ has a single point. In summary, $sp(k)$ encompasses only one single point when $k = 1,3,6,10,15,21,\ldots$, and two points otherwise.

The SCPG sampling set constitutes of symmetrical pairs $\{sp(k)\}_{k\in \mathrm{C}}$ ($\mathrm{C}= \{k_t | t = 1,\ldots,q\}$), instead of data points, which are selected following the Poisson gap distribution rule. Above, $\mathrm{C}$ denotes the selected index set, $k_t$ denotes the $t$-th selected symmetrical pairs index and $q$ denotes the number of selected symmetrical pairs. The gap of neighboring indices, i.e., $g_t = k_{t+1} - k_t - 1$ $(t = 1,\ldots,q-1)$ obeys the Poisson distribution.

If the selected symmetrical pairs consist of two data points, only one point is selected randomly for sampling and then the other non-sampled point is filled with the copy of the sampled one.

***Definition*** 4: In SCPG, if the symmetrical pairs $\{sp(k)\}_{k\in \mathrm{C}}$ ($\mathrm{C}= \{k_t | t = 1,\ldots,q\}$) are selected for sampling, the Symmetrical Point Index Set (SPIS)— $\Omega$ is defined as

$$\begin{aligned}&\Omega=\{(i,j)\}\cup\{(j,i)\},\\&\forall(i,j): 1\le i\le j\le \mathrm{N}, j(j-1)/2+i=k_t, t=1,\dots,q\end{aligned} \tag{13}$$

Let $r$ be the number of elements in $\Omega$. The Filled Sampled Data Vector (FSDV)— $\boldsymbol{y}$ is defined as

$$\boldsymbol{y}=(y_1 \quad y_2 \quad \cdots \quad y_r)^T . \tag{14}$$

where $y_t$ $(t=1,\dots,r)$ is the $t$-th entry of NUS FID data ordered by rows first and then columns. For a vivid demonstration, the corresponding procedures, SPIS and FSDV of SCPG are illustrated with a toy example in Figure 4. As clearly illustrated in Figure 4d, $\boldsymbol{P}_\Omega{}^H\boldsymbol{y}$ is symmetrical, i.e.,

$$\boldsymbol{P}_\Omega{}^H\boldsymbol{y}=\left(\boldsymbol{P}_\Omega{}^H\boldsymbol{y}\right)^T. \tag{15}$$

And $\boldsymbol{P}_\Omega{}^H\boldsymbol{P}_\Omega$ is linear symmetry-invariant operator.

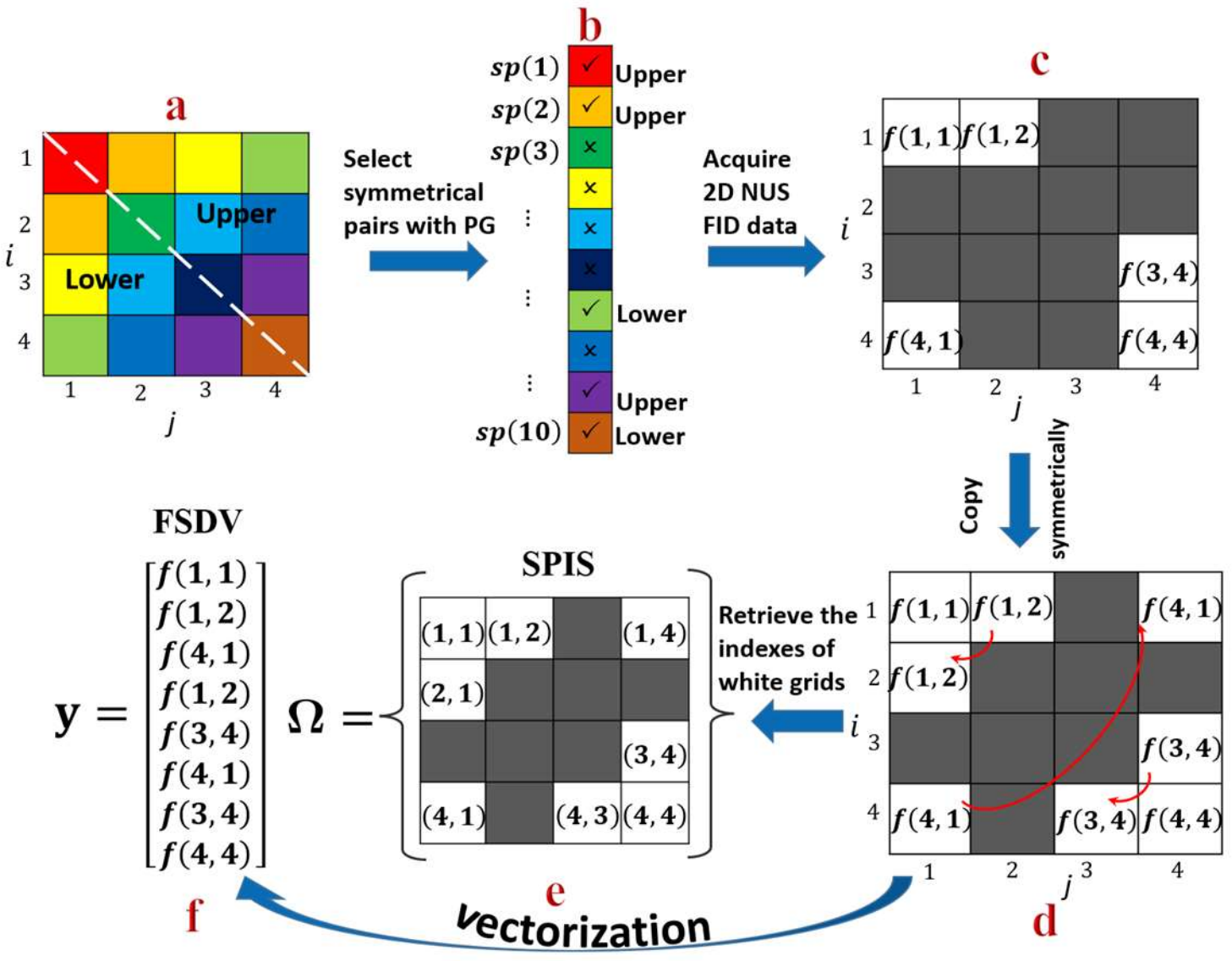

**Figure 4**. Illustration of the procedures, SPIS and FSDV of SCPG with a toy example. (a) The 2D data points indicated by a colored grid table where the grids filled with the same color constitute a symmetrical pair. (b) Five symmetrical pairs are selected from $\{sp(k)\}_{k=1,2,3,\ldots,10}$ with PG. 'Tick' signs indicate the selected symmetrical pairs, while 'cross' signs the ignored symmetrical pairs. At the right side of selected symmetrical pairs, 'Upper' ('Lower') indicate selecting the upper (lower) points of symmetrical pairs for sampling. (c) The grid table indicating the sampled (white) and non-sampled (gray) points. In white grids, $f(i,j)$ denotes the corresponding acquired data. (d) The illustration of symmetrically coping data. (e) The corresponding Symmetrical Point Index Set (SPIS) including the indices of white grids. (f) The corresponding Filled Sampled Data Vector (FSDV) constructed by column-wise vectorization of data points.

## *4.2. SCPG Sampling for 2D NUS CS Reconstruction*

***Theorem*** 2: Let $\Omega$ and $\boldsymbol{y}$ be the SPIS and FSDV of SCPG, respectively; $\Omega_0$ be a half off-diagonal index subset of $\Omega$, ${\Omega_0}^T$ another half off-diagonal index subset of $\Omega$, $\Omega_d$ the diagonal index subset of $\Omega$, i.e., $\Omega = \Omega_0 \cup {\Omega_0}^T \cup \Omega_d$ and $\emptyset = \Omega_0 \cap {\Omega_0}^T \cap \Omega_d$; $\boldsymbol{y}_0$ and $\boldsymbol{y}_d$ be the sub-vector of $\boldsymbol{y}$ extracted according to $\Omega_0$ and $\Omega_d$, respectively. If (16) has unique optimal solution, then two problems (16) and (17) have same optimal solution, i.e., $\boldsymbol{X}_0^* = \boldsymbol{X}_1^*$.

$$\boldsymbol{X}_0^* = \arg\min_{\boldsymbol{X}} \left\|\boldsymbol{P}_\Omega {\boldsymbol{F}_{2\mathrm{D}}}^{-1}\boldsymbol{X} - \boldsymbol{y}\right\|_2^2/2 + \lambda\|\boldsymbol{X}\|_1 \tag{16}$$

$$\boldsymbol{X}_1^* = \arg\min_{\boldsymbol{X}=\boldsymbol{X}^T} \left\|\boldsymbol{P}_{\Omega_0}\boldsymbol{F}_{\mathrm{2D}}{}^{-1}\boldsymbol{X} - \boldsymbol{y}_0\right\|_2^2 + \left\|\boldsymbol{P}_{\Omega_d}\boldsymbol{F}_{\mathrm{2D}}{}^{-1}\boldsymbol{X} - \boldsymbol{y}_d\right\|_2^2/2 + \lambda\|\boldsymbol{X}\|_1 \tag{17}$$

Proof: Adding $\boldsymbol{P}_\Omega{}^H$ into the first term of (16), we have (18) which is the equivalent to (16).

$$\boldsymbol{X}_0^* = \arg\min_{\boldsymbol{X}} \left\|\boldsymbol{P}_\Omega{}^H\boldsymbol{P}_\Omega\boldsymbol{F}_{\mathrm{2D}}{}^{-1}\boldsymbol{X} - \boldsymbol{P}_\Omega{}^H\boldsymbol{y}\right\|_F^2/2 + \lambda\|\boldsymbol{X}\|_1, \tag{18}$$

where $\boldsymbol{P}_\Omega{}^H\boldsymbol{P}_\Omega\boldsymbol{F}_{\mathrm{2D}}{}^{-1}$ is a symmetry-invariant linear operator and $\boldsymbol{P}_\Omega{}^H\boldsymbol{y}$ is a symmetrical matrix. Invoking *Theorem* 1 and the uniqueness of solution to (16), we conclude $\boldsymbol{X}_0^* = \boldsymbol{X}_0^{*T}$. Thus, we have

$$\begin{aligned}\boldsymbol{X}_0^* &= \arg\min_{\boldsymbol{X}=\boldsymbol{X}^T} \left\|\boldsymbol{P}_\Omega\boldsymbol{F}_{\mathrm{2D}}{}^{-1}\boldsymbol{X} - \boldsymbol{y}\right\|_2^2/2 + \lambda\|\boldsymbol{X}\|_1 \\ &= \arg\min_{\boldsymbol{X}=\boldsymbol{X}^T} \left\|\boldsymbol{P}_{\Omega_0}\boldsymbol{F}_{\mathrm{2D}}{}^{-1}\boldsymbol{X} - \boldsymbol{y}_0\right\|_2^2 + \left\|\boldsymbol{P}_{\Omega_d}\boldsymbol{F}_{\mathrm{2D}}{}^{-1}\boldsymbol{X} - \boldsymbol{y}_d\right\|_2^2/2 + \lambda\|\boldsymbol{X}\|_1, \\ &= \boldsymbol{X}_1^*\end{aligned} \tag{19}$$

The uniqueness of solution to (16) is guaranteed if the cardinality of random set $|\Omega|$ is sufficiently large so that the measurement matrix $\boldsymbol{P}_\Omega\boldsymbol{F}_{2\mathrm{D}}{}^{-1}$ satisfies the Restricted Isometry Property (RIP)[39]. As Candès, Romberg, and Tao stated in their paper[31], the solution of (16) is unique with high probability if $|\Omega| \geq C \cdot r \cdot \log(n)$ where $C$ is a known (small) constant, $r$ is the non-zero points number of ideal spectrum, and $n$ is the length of spectrum.

*Theorem* 2 suggests that $l_1$-norm regularized CS reconstruction with SCPG sampling is equivalent to strict symmetrical constraint on regular non-symmetrical sampling data. Thus, the combination of SCPG and CS guides the reconstructed spectrum to be simultaneously sparse and symmetrical. The advantage of (16) over (17) is the friendly compatibility to the existing reconstruction methods, without any extra explicit constraint other than the $l_1$-norm regularization in the original algorithm, which is the main contribution of this work.

So far, all the CS methods we have tried for solving (16) can reconstruct sparse and symmetrical spectra. We are also interested in exploring if the symmetrical property can be guaranteed when 2D IST-D is applied for solving (20):

$$\arg\min_{\boldsymbol{X}} \|\boldsymbol{X}\|_1,\ s.t. \left\|\boldsymbol{P}_\Omega\boldsymbol{F}_{\mathrm{2D}}{}^{-1}\boldsymbol{X} - \boldsymbol{y}\right\|_2 \leq \varepsilon\ . \tag{20}$$

Here, we expand the 1D IST-D into the 2D version, summarized in Table 1, by replacing 1D IFT and FT therein with the 2D counterparts. Obviously, if the optimum of (20) is unique, the optimum of (20) is symmetrical since (20) can be equivalently converted to (16) with a certain value of λ. In what follows, we explore the internal iterative characteristic of 2D IST-D with SCPG.

**Table 1**. The summary of 2D IST-D algorithm

**Input:** $\boldsymbol{y}$, $\Omega$, $\varepsilon$, $maxt$

**Initialize:** $t = 0$, $\boldsymbol{S}^0 = \boldsymbol{F}_{2\mathrm{D}}\boldsymbol{P}_\Omega{}^H\boldsymbol{y}$ , $\boldsymbol{X}^0 = \boldsymbol{0}$ , $\boldsymbol{r}^0 = \boldsymbol{y}$

**while** $t \le maxt$ and $\|\boldsymbol{r}^t\|_2 > \varepsilon$

1. $\beta = 0.99 * \max(|\boldsymbol{S}^t|) * (maxt - t)/maxt$
2. $\boldsymbol{X}^{t+1} = \boldsymbol{X}^t + \mathrm{SHR}_\beta(\boldsymbol{S}^t)$
3. $\boldsymbol{r}^{t+1} = \boldsymbol{y} - \boldsymbol{P}_\Omega\boldsymbol{F}_{2\mathrm{D}}{}^{-1}\boldsymbol{X}^{t+1}$
4. $\boldsymbol{S}^{t+1} = \boldsymbol{F}_{2\mathrm{D}}\boldsymbol{P}_\Omega{}^H\boldsymbol{r}^{t+1}$
5. $t = t + 1$

**end**

**Output**: $\boldsymbol{X}^t$ as the result.

***Lemma*** 4: If $\Omega$ and $\boldsymbol{y}$ are SPIS and FSDV of SCPG, respectively; $\boldsymbol{X}^0$ is initialized to be symmetrical, i.e., $\boldsymbol{X}^0 = (\boldsymbol{X}^0)^T$; and $\boldsymbol{X}^{t+1} = \mathrm{SHR}_\beta\left(\boldsymbol{F}_{2\mathrm{D}}\boldsymbol{P}_\Omega{}^H\left(\boldsymbol{P}_\Omega\boldsymbol{F}_{2\mathrm{D}}{}^{-1}\boldsymbol{X}^t - \boldsymbol{y}\right)\right)$, then $\boldsymbol{X}^t = (\boldsymbol{X}^t)^T$ for any $t \ge 1$.

Proof: Combined with *Theorem* 2, we have

$$\boldsymbol{F}_{2\mathrm{D}}\boldsymbol{P}_\Omega{}^H\boldsymbol{y} = \left(\boldsymbol{F}_{2\mathrm{D}}\boldsymbol{P}_\Omega{}^H\boldsymbol{y}\right)^T. \tag{21}$$

If a square matrix $\boldsymbol{X}$ is symmetrical, we have

$$\left(\boldsymbol{F}_{2\mathrm{D}}\boldsymbol{P}_\Omega{}^H\boldsymbol{P}_\Omega\boldsymbol{F}_{2\mathrm{D}}{}^{-1}\boldsymbol{X}\right)^T = \boldsymbol{F}_{2\mathrm{D}}\boldsymbol{P}_\Omega{}^H\boldsymbol{P}_\Omega\boldsymbol{F}_{2\mathrm{D}}{}^{-1}\boldsymbol{X}, \tag{22}$$

which can be easily proved. Through *Theorem* 2, we have $\left(\boldsymbol{F}_{2\mathrm{D}}{}^{-1}\boldsymbol{X}\right)^T = \boldsymbol{F}_{2\mathrm{D}}{}^{-1}\boldsymbol{X}$ and $(\boldsymbol{F}_{2\mathrm{D}}\boldsymbol{X})^T = \boldsymbol{F}_{2\mathrm{D}}\boldsymbol{X}$ for any symmetrical square matrix $\boldsymbol{X}$. Additionally, we have $\left(\boldsymbol{P}_\Omega{}^H\boldsymbol{P}_\Omega\boldsymbol{X}\right)^T = \boldsymbol{P}_\Omega{}^H\boldsymbol{P}_\Omega\boldsymbol{X}$ for any symmetrical square matrix $\boldsymbol{X}$ and thus the equality (22). Through (21) and (22), we conclude that $\mathrm{SHR}_\beta\left(\boldsymbol{F}_{2\mathrm{D}}\boldsymbol{P}_\Omega{}^H\left(\boldsymbol{P}_\Omega\boldsymbol{F}_{2\mathrm{D}}{}^{-1}\boldsymbol{X}^t -\right.\right.$

$\boldsymbol{y}))$ is symmetrical if $\boldsymbol{X}^t$ is symmetrical. Therefore, if $\boldsymbol{X}^0 = (\boldsymbol{X}^0)^T$, then we have $\boldsymbol{X}^t = (\boldsymbol{X}^t)^T$ for any $t \geq 1$.

***Theorem*** 3: If $\Omega$ and $\mathbf{y}$ are the SPIS and FSDV of SCPG, respectively, and $\{\boldsymbol{X}^t\}$ and $\{\boldsymbol{S}^t\}$ are generated by the 2D IST-D algorithm in Table 1, then $\boldsymbol{X}^t$ and $\boldsymbol{S}^t$ are symmetrical, i.e., $\boldsymbol{X}^t = (\boldsymbol{X}^t)^T$ and $\boldsymbol{S}^t = (\boldsymbol{S}^t)^T$, for any $t \geq 0$.

Proof: We use mathematical induction method to prove *Theorem* 3. Assuming that $\boldsymbol{X}^t = (\boldsymbol{X}^t)^T$, we have

$$\begin{aligned}(\boldsymbol{X}^{t+1})^T &= (\boldsymbol{X}^t)^T + \left(\mathrm{SHR}_\beta\left(\boldsymbol{F}_{2\mathrm{D}}\boldsymbol{P}_\Omega{}^H\boldsymbol{y} - \boldsymbol{F}_{2\mathrm{D}}\boldsymbol{P}_\Omega{}^H\boldsymbol{P}_\Omega\boldsymbol{F}_{2\mathrm{D}}{}^{-1}\boldsymbol{X}^t\right)\right)^T \\ &= \boldsymbol{X}^t + \mathrm{SHR}_\beta\left(\boldsymbol{F}_{2\mathrm{D}}\boldsymbol{P}_\Omega{}^H\boldsymbol{y} - \boldsymbol{F}_{2\mathrm{D}}\boldsymbol{P}_\Omega{}^H\boldsymbol{P}_\Omega\boldsymbol{F}_{2\mathrm{D}}{}^{-1}\boldsymbol{X}^t\right) \\ &= \boldsymbol{X}^{t+1}\end{aligned} \quad . \tag{23}$$

The second equality in (25) holds due to ***Lemma*** 4. Owing to $\boldsymbol{X}^0 = (\boldsymbol{X}^0)^T = \boldsymbol{0}$, we conclude that $\boldsymbol{X}^t = (\boldsymbol{X}^t)^T$ for any $t \geq 0$.

Since

$$\begin{aligned}\boldsymbol{S}^t &= \boldsymbol{F}_{2\mathrm{D}}\boldsymbol{P}_\Omega{}^H\boldsymbol{r}^t = \boldsymbol{F}_{2\mathrm{D}}\boldsymbol{P}_\Omega{}^H\left(\boldsymbol{y} - \boldsymbol{P}_\Omega\boldsymbol{F}_{2\mathrm{D}}{}^{-1}\boldsymbol{X}^t\right) \\ &= \boldsymbol{F}_{2\mathrm{D}}\boldsymbol{P}_\Omega{}^H\boldsymbol{y} - \boldsymbol{F}_{2\mathrm{D}}\boldsymbol{P}_\Omega{}^H\boldsymbol{P}_\Omega\boldsymbol{F}_{2\mathrm{D}}{}^{-1}\boldsymbol{X}^t\end{aligned} \quad , \tag{24}$$

we have

$$\begin{aligned}(\boldsymbol{S}^t)^T &= \left(\boldsymbol{F}_{2\mathrm{D}}\boldsymbol{P}_\Omega{}^H\boldsymbol{y} - \boldsymbol{F}_{2\mathrm{D}}\boldsymbol{P}_\Omega{}^H\boldsymbol{P}_\Omega\boldsymbol{F}_{2\mathrm{D}}{}^{-1}\boldsymbol{X}^t\right)^T \\ &= \left(\boldsymbol{F}_{2\mathrm{D}}\boldsymbol{P}_\Omega{}^H\boldsymbol{y}\right)^T - \boldsymbol{F}_{2\mathrm{D}}\boldsymbol{P}_\Omega{}^H\boldsymbol{P}_\Omega\boldsymbol{F}_{2\mathrm{D}}{}^{-1}(\boldsymbol{X}^t)^T \\ &= \boldsymbol{F}_{2\mathrm{D}}\boldsymbol{P}_\Omega{}^H\boldsymbol{y} - \boldsymbol{F}_{2\mathrm{D}}\boldsymbol{P}_\Omega{}^H\boldsymbol{P}_\Omega\boldsymbol{F}_{2\mathrm{D}}{}^{-1}\boldsymbol{X}^t \\ &= \boldsymbol{S}^t\end{aligned} \quad . \tag{25}$$

Through *Theorem* 3, we show that 2D IST-D is restricted within the symmetrical subspace to search the optimal sparse result during iterations, which, to some extent, improves the searching effectiveness. On the other hand, 2D IST-D can always find a symmetrical solution even if the solution of (20) is not unique.

## 5. Results and Discussion

In this section, we use simulated and experimental data to verify the effectiveness of SCPG by comparing the results of SCPG with those of Random sampling and 2D Woven PG. For sufficient verification of the powerfulness and robustness of SCPG in CS methods, all results are obtained from the CS greedy algorithm— 2D IST-D and a recent CS $l_1$-norm regularized optimization algorithm— Sparse Complex-valued REconstruction Enabled by Newton method (SCREEN)[40]. In SCREEN, a second order optimization algorithm— Truncated Newton Interior Point Method (TNIPM) is applied[41], which is expected to provide faster and more accurate convergence than first order optimization algorithms, e.g., ISTA and FISTA[42]. Regularization parameters of the $l_1$-norm regularized model are uniformly set as 0.001 in this work.

**Synthetic data.** All fully sampled synthetic data are 256×256 and consist of 25 diagonal peaks and 50 pairs (100 peaks) of cross peaks. The 2D synthetic data can be represented by

$$\boldsymbol{W}(t_1,t_2) = \boldsymbol{D}(t_1,t_2) + \boldsymbol{C}(t_1,t_2) + \boldsymbol{C}(t_2,t_1), \tag{26}$$

where $t_1 = 1,2,3,\dots,256$, $t_2 = 1,2,3,\dots,256$. $\boldsymbol{D}(t_1,t_2)$ is the signal of 25 diagonal peaks, formulated as

$$\boldsymbol{D}(t_1,t_2) = \sum_{n=1}^{25} d_n \exp(2\pi j f_n t_1 - \alpha t_1)\exp(2\pi j f_n t_2 - \alpha t_2), \tag{27}$$

where $d_n$ is the amplitude of the $n$-th diagonal peak which is randomly set within 9~10; Frequencies of 25 diagonal peaks, $(f_n, f_n), n = 1,2,3,\dots,25$, are uniformly randomly chosen from the discrete frequency set $\{(N/256, N/256)|N = 0,1,2,\dots,255\}$; $\alpha$ is the decay constant which is set as 0.001 both for all diagonal and cross peaks. $\boldsymbol{C}(t_1,t_2) + \boldsymbol{C}(t_2,t_1)$ is the signal of 50 pairs of cross peaks and $\boldsymbol{C}(t_1,t_2)$ is formulated as

$$\boldsymbol{C}(t_1,t_2) = \sum_{m=1}^{50} c_m \exp(2\pi j f_{m1} t_1 - \alpha t_1)\exp(2\pi j f_{m2} t_2 - \alpha t_2), \tag{28}$$

where $c_m$ is the amplitude of the $m$-th pair of cross peak which is uniformly randomly set in the range of 3~4; Frequencies of 50 pairs of cross peaks, $(f_m^1, f_m^2), m = 1,2,3,\dots,50$, are randomly chosen from the set of discrete frequency pairs $\{((N/256, M/256)|N = 0,1,2,\dots,255; M = 0,1,2,\dots,255;\ N \neq M\}$. The statistical analysis of reconstructions of synthetic data is based on Monte Carlo simulation. 100 fully sampled synthetic data, $\{\boldsymbol{W}_i(t_1,t_2)|i = 1,2,3,\dots,100\}$, are stochastically generated according to (26). Then, $\boldsymbol{W}_i(t_1,t_2)$ is scaled by its maximum absolute value to obtain the scaled data

$$\overline{\boldsymbol{W}}_i(t_1,t_2) = \boldsymbol{W}_i(t_1,t_2)/\max\left(\text{abs}\left(\boldsymbol{W}_i(t_1,t_2)\right)\right). \tag{29}$$

For $i = 1,2,3,\dots,100$, $\overline{\boldsymbol{W}}_i(t_1,t_2)$ is added with 10 different levels (measured with standard deviations) of Gaussian white noise. The Gaussian white noise is with the mean 0 and the standard deviations 0 (non-noisy), 2.5, 5, 7.5, 10, 12.5, 15, 17.5, 20, and 22.5 ($\times10^{-4}$), respectively. Then, for each data of every noisy level, 10 NUS rates (nusr), 5%, 7.5%, 10%, 12.5%, 15%, 17.5%, 20%, 22.5%, 25%, and 27.5% are applied according to Random, 2D Woven PG, and SCPG sampling schedules, respectively. Through the foregoing preparation, we have $100\times10\times10\times3$=30000 synthetic NUS data. We use IST-D and SCREEN to reconstruct these 30000 synthetic NUS data for statistical analysis (See Figure 5). The reconstruction error is evaluated by the Relative $l_2$-norm Error (RLNE), which is defined as

$$\epsilon = \|\dot{\boldsymbol{x}} - \acute{\boldsymbol{x}}\|_2 / \|\dot{\boldsymbol{x}}\|_2, \tag{30}$$

where $\epsilon$ is RLNE, and $\dot{\boldsymbol{x}}$ and $\acute{\boldsymbol{x}}$ denotes the vector consisting of true and reconstructed amplitudes of spectral peaks, respectively. The smaller RLNE is, the more accurate the peak recovery is. Here, considering the amplitudes of diagonal peaks are significantly greater than those of cross peaks, the RLNE of whole spectral peaks would dominantly depend on the diagonal peaks. Thus, we separately discuss the reconstructions of diagonal and cross peaks by letting $\dot{\boldsymbol{x}}$ and $\acute{\boldsymbol{x}}$ only encompass amplitudes of diagonal or cross peaks.

The top panel in Figure 5 displays the statistical results of diagonal peaks, while the bottom one shows those of cross peaks. The statistical results, mean and standard deviation (STD), are all presented with colored images consisting of $10\times10$ pixels, each representing a certain noise level and NUS rate. The RLNE values are indicated by the colors according to the color bars on the right. For convenient comparison, the color bars are the same for all statistical results of diagonal peaks (Figure 5 a1~a3, b1~b3, f1~f3, and g1~g3) and the same for all statistical results of cross peaks (Figure 5 c1~c3, d1~d3, h1~h3, and i1~i3). For a clear comparison, the boundaries of grids representing values above and below 1/3 of the value range are marked with black lines. Figure 5 a1~a3 and b1~b3 (c1~c3 and d1~d3) show the mean (standard deviation) values of RLNE of diagonal peaks reconstructed by IST-D and SCREEN, respectively. We can observe that the three sampling schedules perform almost equally on the reconstruction of diagonal peaks. In fact, all RLNE values of diagonal peaks are almost less than 0.05, which is satisfactorily accurate. In Figure 5 e1 and e2, we demonstrate all 100 RLNE of diagonal peaks reconstructed by IST-D and SCREEN in nusr = 5% and noise STD = 0.0010. As we can see, in both IST-D and SCREEN reconstructions, the RLNE lines of diagonal peaks of three sampling schedules intertwine one another, which reveals that the three sampling schedules do not differ apparently in the reconstruction performance of diagonal peaks.

For the cross peak reconstruction, we infer from Figure 5 f1~f3 and g1~g3 that 2D Woven PG sampling does improve the reconstruction of cross peaks compared to Random sampling, and SCPG sampling further improves the reconstruction of cross peaks relative to 2D Woven PG sampling. With the 1/3 boundary in Figure 5 f1~f3, g1~g3, h1~h3 and i1~i3, we could clearly identify that the areas of 1/3 RLNE range of SCPG, reconstructed by IST-D or SCREEN, are less than 2D Woven PG and Random sampling, indicating that SCPG is more efficient and robust than 2D Woven PG and Random sampling. Figure 5 k1 and k2 demonstrate all 100 RLNE of cross peaks reconstructed by IST-D and SCREEN in nusr = 5% and noise STD = 0.0010. Therein, three RLNE lines of three sampling schedules are clearly separated. The RLNE lines of SCPG are at the bottom, those of 2D Woven PG in the middle, and those of Random sampling at the top, which substantiates the powerfulness and robustness of SCPG sampling over the other two sampling methods.

**Experimental Data**. Here, we use two experimental data, originally downloaded from [37], to validate the performance of SCPG sampling in practical biochemical applications. The first data is a 512×512 2D symmetrical spectra data obtained from the 2D $^{13}C$-$^{13}C$ projection of 4D methyl HMQC-NOESY-HMQC spectrum collected at 600 MHz $^{1}H$ frequency, 200 ms NOE mixing time, for a 0.5 mM sample of the ILV methyl-labeled m04 protein of cytomegalovirus[43]. The second one is a 620×620 2D symmetrical spectra data obtained from $^{15}N$-$^{15}N$ projection of 3D (H)N(COCO)NH spectrum of a-synuclein, carried out on a 600 MHz NMR equipment[36]. Based on Monte Carlo simulation, we stochastically generate 100 Random, 2D Woven PG, and SCPG sampling schedules, respectively, according to which 5% data point subsets of the two data are chosen to generate the 5% NUS NMR data.

The traditional reconstruction is implemented slice by slice along the direct dimension and then with a sum up of all the reconstructed slices to generate the projection spectrum. Apparently, slice reconstruction is very time-consuming, and the slice spectrum is not necessarily symmetrical. Here, IST-D and SCREEN are employed to directly reconstruct the whole 2D projection spectrum from the sum of NUS FID signals, which maintains the symmetrical structure and is time-saving compared to the slice-wise reconstruction. Considering that the projection spectrum is overwhelmingly more sophisticated than the slice spectrum, the direct reconstruction is more challenging than the slice-wise reconstruction and thus relies more on the efficient sampling schedule to retrieve the desired spectral information.

The fully sampled spectra serve as the references for the evaluation of peak recovery. The recovery evaluations of IST-D and SCREEN are both carried out based on multiple indicators— RLNE, correlation coefficient of intensities of fully sampled peaks and reconstructed peaks ('R'), linear fitting coefficients of intensities of fully sampled peaks and reconstructed peaks (i.e., the slope 'a' and intercept 'b'), and reconstruction time.

As listed in Table 2, in each rectangle grid which encompasses a group of means and standard deviations (MEAN$\pm$STD) of 100 Monte Carlo trials, the same indicator is evaluated on the recovery of diagonal and cross peaks separately. The value indicating the most excellent or fastest reconstruction among the three sampling schedules is in bold font. Among the three sampling schedules, Random sampling is the worst for peak recovery, which reveals that Random sampling does not work well at such a low NUS rate; 2D Woven PG sampling obviously improves the peak recovery compared to Random sampling; SCPG is the most excellent and robust sampling schedule for the most accurate recovery of peaks and the least reconstruction time (i.e., accelerate the convergence of algorithms). The fast reconstruction resulting from SCPG can be attributed to two reasons: first, PG sampling in symmetrical pairs is more effective for retrieving essential spectral information; second, symmetrically copying reduces the original non-sampled data points that need to be reconstructed.

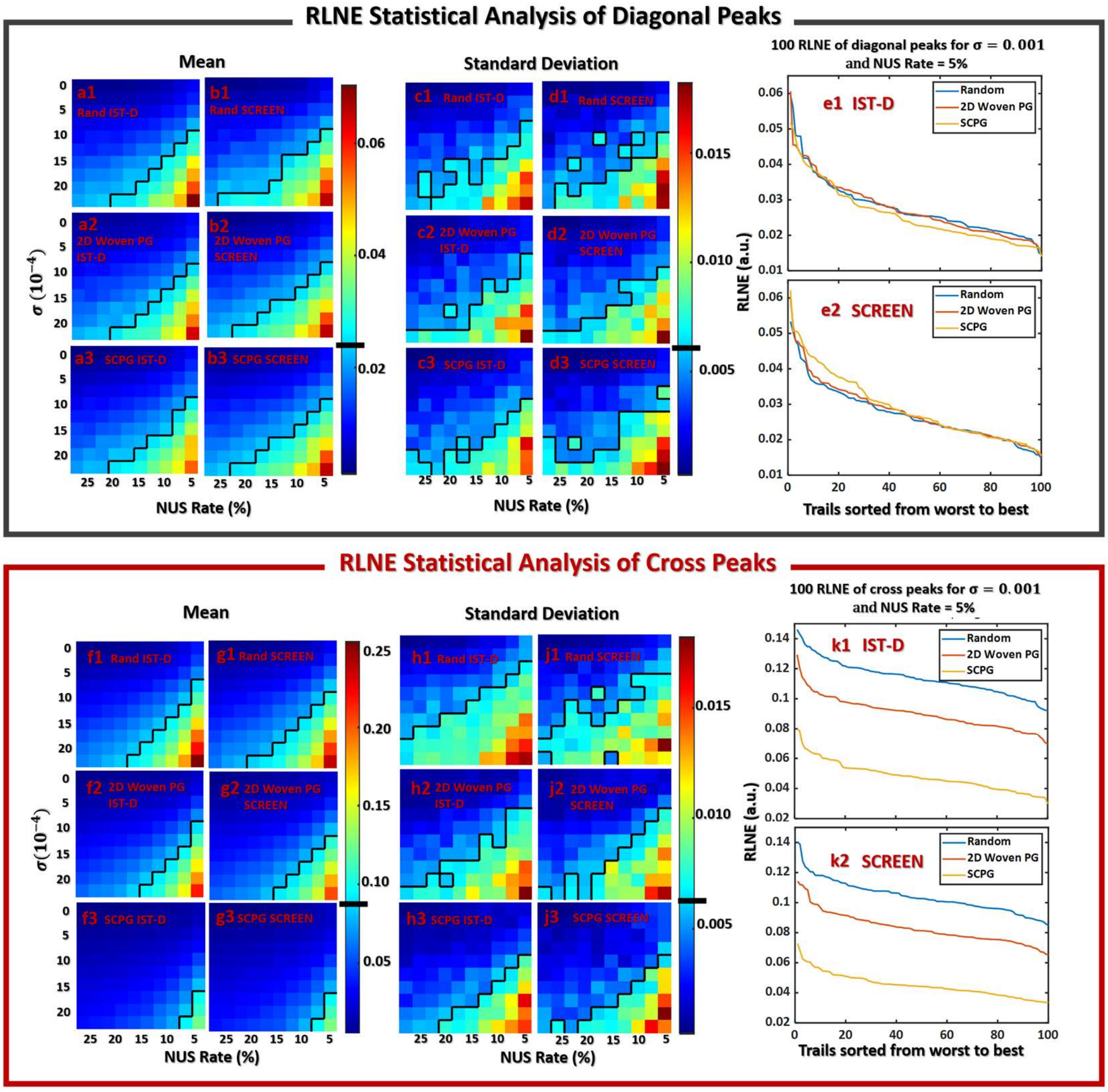

**Figure 5** The statistical analysis of reconstructions of synthetic data based on Monte Carlo simulation. The top (bottom) part shows the statistic of RLNE of diagonal (cross) peaks. a1~a3 and b1~b3 (c1~c3 and d1~d3) show the means (Standard deviation, STD) of RLNE of diagonal (cross) peaks in combinations of 10 NUS rates and 10 noisy levels with three sampling schedules and reconstructed by IST-D and SCREEN, respectively. 100 descending-sorted RLNE in NUS Rate 5% and noise level σ=0.001 is selected and shown in e1 and e2. f1~f3, g1~g3, h1~h3, j1~j3 and k1~k2 correspond to a1~a3, b1~b3, c1~c3, d1~d3 and e1~e2 but the content is about cross peaks, instead. The grid boundary between the values greater and less than 1/3 of the whole value range is marked with black lines in a1~a3, b1~b3, c1~c3, d1~d3, f1~f3, g1~g3, h1~h3 and j1~j3, as indicated in the right color bars.

**Table 2**. Statistical estimation (MEAN±STD) of peak recovery of 2 experimental 5% NUS symmetrical NMR data by IST-D and SCREEN through 100 random, 100 2D Woven PG (2D WPG) and 100 SCPG sampling schedules which are generated based on Monte Carlo simulation.

| Data | Algorithms | Sampling | Peak Type | RLNE | a | b | R | Time (s) |
|---|---|---|---|---|---|---|---|---|
| $^{13}C$-$^{13}C$ NMR Symmetrical Spectrum | IST-D | Random | Diagonal | 0.098±0.023 | 1.01±0.02 | -0.66±0.27 | 0.9963±0.0014 | 21±3 |
| | | | Cross | 0.706±0.040 | 0.43±0.06 | -0.18±0.03 | 0.8876±0.0294 | |
| | | 2D WPG | Diagonal | 0.029±0.008 | 1.00±0.01 | -0.15±0.04 | 0.9995±0.0002 | **21±1** |
| | | | Cross | 0.130±0.016 | 0.95±0.02 | -0.08±0.01 | 0.9939±0.0014 | |
| | | Our SCPG | Diagonal | **0.012±0.004** | **1.01±0.01** | **-0.02±0.01** | **0.9999±0.0000** | **21±1** |
| | | | Cross | **0.040±0.005** | **0.99±0.01** | **-0.03± 0.01** | **0.9994±0.0002** | |
| | SCREEN | Random | Diagonal | 0.082±0.023 | 1.00±0.01 | -0.47±0.21 | 0.9976±0.0012 | 83±13 |
| | | | Cross | 0.647±0.043 | 0.50±0.06 | -0.20±0.03 | 0.9114±0.0235 | |
| | | 2D WPG | Diagonal | 0.021±0.005 | 1.01±0.00 | -0.13±0.03 | 0.9997±0.0001 | 61±3 |
| | | | Cross | 0.126±0.013 | 0.95±0.02 | -0.08±0.01 | 0.9950±0.0011 | |
| | | Our SCPG | Diagonal | **0.007±0.002** | **1.00±0.00** | **-0.02±0.01** | **1.0000±0.0000** | **39±1** |
| | | | Cross | **0.042±0.005** | **0.98±0.01** | **-0.03±0.01** | **0.9996±0.0001** | |
| $^{15}N$-$^{15}N$ NMR Symmetrical Spectrum | IST-D | Random | Diagonal | 0.113±0.027 | 0.96±0.01 | -0.26±0.07 | 0.9948±0.0023 | 62±2 |
| | | | Cross | 0.798±0.024 | 0.30±0.04 | -0.08±0.01 | 0.7162±0.0299 | |
| | | 2D WPG | Diagonal | 0.042±0.007 | 0.98±0.01 | -0.08±0.02 | 0.9993±0.0002 | 65±2 |
| | | | Cross | 0.230±0.011 | 0.89±0.02 | -0.04±0.01 | 0.9785±0.0030 | |
| | | Our SCPG | Diagonal | **0.010±0.002** | **1.00±0.01** | **-0.01±0.01** | **0.9999±0.0000** | **52±1** |
| | | | Cross | **0.090±0.005** | **0.95±0.01** | **-0.01±0.00** | **0.9971±0.0005** | |
| | SCREEN | Random | Diagonal | 0.088±0.019 | 0.97±0.01 | -0.20±0.05 | 0.9968±0.0013 | 151±33 |
| | | | Cross | 0.737±0.024 | 0.40±0.04 | -0.05±0.01 | 0.7776±0.0267 | |
| | | 2D WPG | Diagonal | 0.038±0.006 | 0.99±0.01 | -0.07±0.01 | 0.9994±0.0001 | 134±6 |
| | | | Cross | 0.218±0.010 | 0.89±0.02 | -0.04±0.00 | 0.9843±0.0023 | |
| | | Our SCPG | Diagonal | **0.018±0.002** | **1.00±0.00** | **-0.03±0.00** | **0.9996±0.0000** | **97±4** |
| | | | Cross | **0.086±0.004** | **0.95±0.01** | **-0.01±0.00** | **0.9978±0.0004** | |

Note: The means and standard deviations are computed over all 100 tests, respectively. In the grids encompassing a group of MEAN±STD, the comparison is implemented for diagonal and cross peaks separately. Therein, the values of 'RLNE' whose means is the smallest, the values of 'a' whose means are most closed to 1, the values of 'b' whose means are most closed to 0, the values of 'R' whose means are most closed to 1 and the reconstruction time whose means is the least are in bold font. If more than one means are equal, the value with the smallest standard deviation is in bold font (Although, it does not happen in this table).

Additionally, some interesting conclusions can be drawn according to the statistical comparisons in Table 2. For instance, the recovery of diagonal peaks is always more accurate than that of cross peaks. It is because diagonal peaks are more insensitive to NUS artifact and experimental noise than cross peaks, benefitting from their multiple times higher intensities over cross peaks. The reconstructions of IST-D are faster than those of SCREEN since the variation of thresholding value in IST- would speed up the convergence.

Figure 6 shows the reconstructed results of 5% NUS $^{13}C$-$^{13}C$ data of the three sampling schedules by IST-D (top panel) and SCREEN (bottom panel). We sort the 100 groups of results of $^{13}C$-$^{13}C$ reconstruction in ascending order according to the RLNE values of cross peaks and choose the 50$^{th}$ reconstructed results to show. In the top panel, as shown in Figure 6 a1~a4, diagonal peaks are satisfactorily recovered through all three sampling schedules, but the recovery of cross peaks differs among the three sampling methods. Compared to the fully sampled spectrum (Figure 6 a1), almost all cross peaks are absent in the reconstructed spectrum sampled through Random sampling schedule (Figure 6 a2); conspicuous artifacts are presented, and a number of cross peaks disappear in the reconstructed spectrum sampled through 2D Woven PG sampling schedule (Figure 6 a3). In contrast, in the reconstructed spectrum sampled through SCPG sampling schedule (Figure 6 a3), artifacts are evidently removed, and cross peaks are relatively well recovered. To simplify the presentation, we do not show the RLNE of diagonal peaks but only that of the cross peaks in $^{13}C$-$^{13}C$ data (Figure 6 b1~b4, d1~d3).

For a clear comparison of the recovery of cross peaks, we use a homemade MATLAB script to locate cross peaks and calculate their integral intensities, which are shown in ascending order in Figure 6 b1~b4. It can be clearly observed that a majority of cross peaks are absent in the reconstructed spectrum sampled through Random sampling schedule (Figure 6 b2, RLNE=0.716); cross peaks are weakened and even disappear in the reconstructed spectrum sampled through 2D Woven PG sampling schedule (Figure 6 b3, RLNE=0.125). On the contrary, cross peaks are well recovered and appear in pairs in the reconstructed spectrum sampled through 2D SCPG sampling schedule (Figure 6 b4, RLNE=0.039). In Figure 6 b4, peak intensities of a pair are equal, verifying that SCPG sampling schedule enforces CS methods to use the symmetrical structure of NMR spectrum and reconstruct an ideally symmetrical spectrum. In the bottom panel, as shown in Figure 6 c~d, the reconstructed results by SCREEN through the three sampling schedules are consistent with the analysis of IST-D above, which corroborates that SCPG maintains the powerfulness in both CS greedy and $l_1$-norm regularized methods.

**6. Conclusion**

The symmetrical NMR spectroscopy is an effective tool for biochemical analysis but suffers from the time-consuming acquisition process. NUS serves as a common strategy for accelerating acquisition, but the symmetrical NMR NUS reconstruction is of poor quality, particularly for cross peaks that are several orders of magnitude weaker than diagonal peaks. For improving the reconstruction quality of cross peaks, based on CS methods, we explore the exclusive property of the symmetrical NMR spectrum and prove that the symmetrical constraint can significantly improve the recovery of cross peaks. Rather than explicitly imposing the symmetrical constraint in the reconstruction model, we implicitly constrain the reconstructed spectrum to be symmetrical with the proposed NUS sampling schedule— SCPG, where symmetrical pairs are selected with 1D PG and sampled data points are copied to their symmetrical peers. We theoretically prove the effectiveness of the symmetrical constraint of SCPG, and the statistical analysis of substantial reconstructions of simulated and experimental data substantiates that SCPG is apparently more effective and robust than the state-of-art sampling schedule (2D Woven PG) for NUS symmetrical NMR reconstruction. Although SCPG is proposed for NUS symmetrical NMR reconstruction, it has great potential to generalize to other recovery problems of data with symmetrical structures.

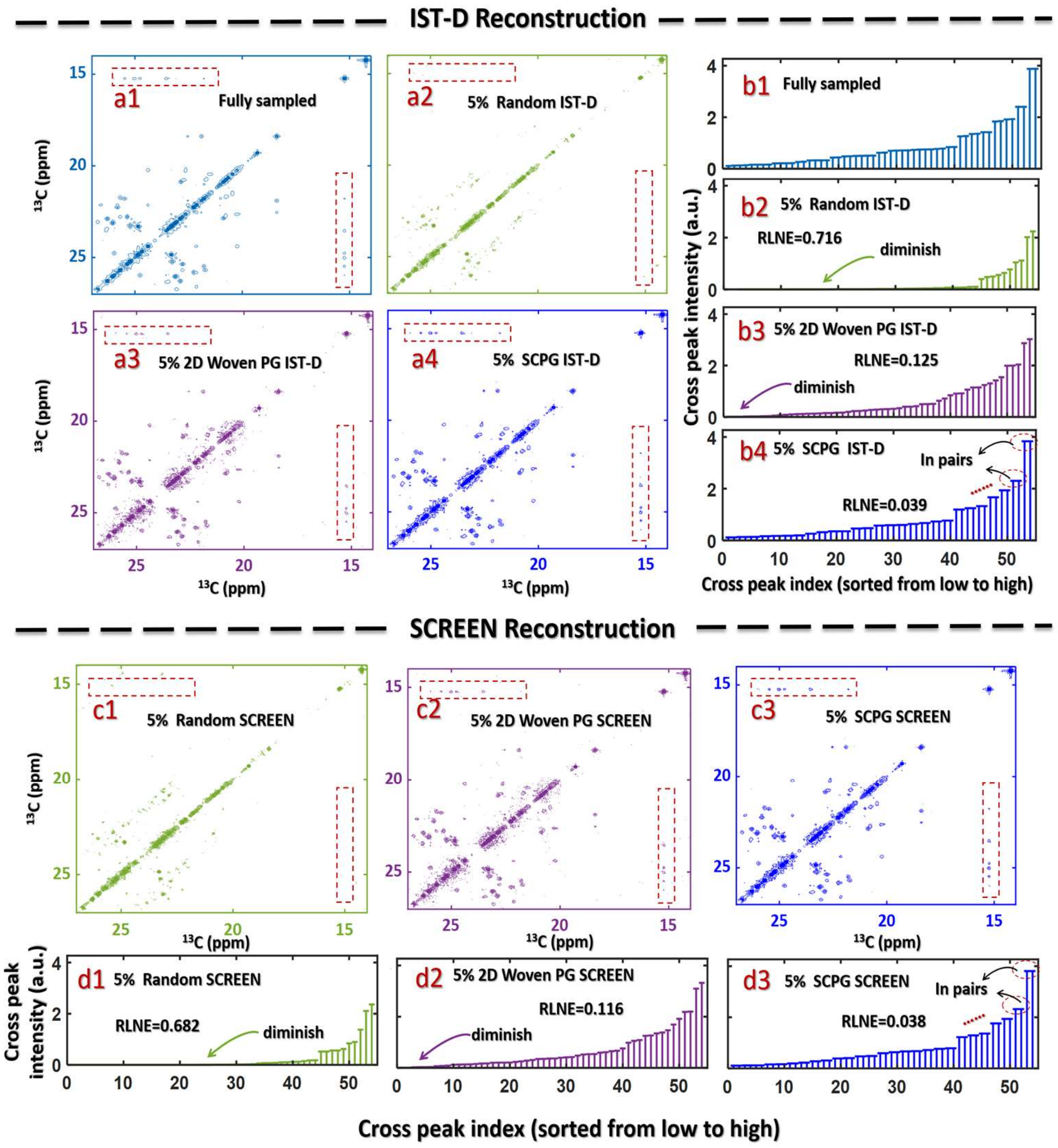

**Figure 6.** Results of the 50th reconstruction of NUS $^{13}C$-$^{13}C$ data. a1 is the fully sampled spectrum and a2~a4 are reconstructed spectra by IST-D with 5% Random, 2D Woven PG and SCPG sampling, respectively. Some areas with weak cross peaks are circled with dash red rectangles for comparison. b1 shows the ascending-sorted integral intensities of cross peaks in fully sampled spectrum. b2~b4 show the ascending-sorted integral intensities of cross peaks recovered by IST-D through Random, 2D Woven PG and SCPG sampling, respectively. The related RLNE and obviously diminished cross peaks are marked in b2 and b3. In b4, cross peaks are presented in pairs as indicated by dash red circles. c1~c3 and d1~d3 correspond to a2~a4 and b2~b4 but the results are obtained by SCREEN, instead.

**Declaration of Competing Interest**

The authors declare that they have no known competing financial interests or personal relationships that could have appeared to influence the work reported in this paper.

**CRediT authorship contribution statement**

**Enping Lin**: Conceptualization, Methodology, Software, Writing - Original Draft. **Ze Fang**: Methodology, Validation, Writing - Original Draft. **Yuqing Huang**: Validation. **Yu Yang**: Supervision, Writing - Review & Editing. **Zhong Chen**: Supervision, Writing – Review & Editing.

**Acknowledgements**

We thank Dr. Sven Hyberts from Harvard Medical School for sharing the C source code of Poisson Gap Schedule. This work was partially supported by the National Natural Science Foundation of China U1805261.